\documentclass[11pt,twoside]{article}

\usepackage{asp2014}

\aspSuppressVolSlug
\resetcounters

\begin{document}

\title{The Research Data Alliance: Building Bridges to Enable Scientific Data Sharing}

% Note the position of the comma between the author name and the 
% affiliation number.
% Author names should be separated by commas.
% The final author should be preceded by "and".
% Affiliations should not be repeated across multiple \affil commands. If several
% authors share an affiliation this should be in a single \affil which can then
% be referenced for several author names.
% See ManuscriptInstructions.pdf and ASPmanual2010.pdf 3.1.4 for more details
\author{Fran\c{c}oise Genova,
\affil{Observatoire astronomique de Strasbourg, Universit\'{e} de Strasbourg, CNRS, UMR 7550, Strasbourg, France}
}

% This section is for ADS Processing.  There must be one line per author.
\paperauthor{Fran\c{c}oise Genova}{francoise.genova@astro.unistra.fr}{0000-0002-6318-5028}{Observatoire astronomique de Strasbourg, Universit\'{e} de Strasbourg, CNRS, UMR 7550}{CDS}{Strasbourg}{N/A}{F-67000}{France} 

\begin{abstract}
The Research Data Alliance \footnote{https://rd-alliance.org/node} is an international organization which aims at building the technical and sociological bridges that enable the open sharing of scientific data. It is a remarkable forum to discuss all the aspects of scientific data sharing with colleagues from all around the world: in November 2016, after 3.5 years of existence, it has 4 500 members from 115 countries. The biannual Plenary meetings, which gather several hundred participants, are rotating between different regions. The March 2017 one will be held in Barcelona and the September 2017 one in Montreal, after Tokyo and Denver in 2016. The RDA work is organized bottom-up, with Working Groups which have 18 months to produce \textit{implementable} deliverables and Interest Groups which serve as platforms of communication and discussion and also produce important outputs such as surveys and recommendations. There are currently 27 Working Groups and 45 Interest Groups, tackling a wide diversity of subjects, including community needs, reference for sharing, data stewardship and services, and topics related to the base infrastructure of data sharing. Some scientific communities use the RDA as a neutral forum to define their own disciplinary data sharing framework, with major successes such as the Wheat Data Interoperability Working Group which worked in coordination with the International Wheat Initiative. Astronomy has the IVOA to define its interoperability standards, and so we do not need to create a Group for that purpose in the RDA. But many topics discussed in the RDA have a strong interest for us, for instance on data citation or certification of data repositories. We have a lot to share from what we have learnt in building our disciplinary global data infrastructure; we also have a lot to learn from others. The paper discusses RDA current themes or results of interest for astronomy data providers, and current liaisons with astronomy. 
\end{abstract}

\section{Introduction}

The Research Data Alliance and the way its Working Groups and Interest Groups are organised are described in \citet{PO38_adassxxv}. Since last year, the RDA continued to develop at a remarkable pace. In November 2016, it counts more than 4500 members (from about 3000 during the same period last year) from 115 different countries. The Eighth Plenary meeting was held in Denver, 15-17 September 2016, in the framework of the International Data Week which also included the SciDataCon 2016 meeting \textit{Advancing the Frontiers of Data in Research} organised by CODATA and the Wold Data System (WDS). This was once again a major gathering of people from all around the world involved in the sharing of scientific data. All relevant profiles were represented, as they have been since the beginning of the RDA in March 2013. This includes researchers, data centre staff, technology developers, librarians, publishers, but also policy makers and research funders. Scientific data sharing is currently a hot topic at government level in many countries and for many research funding agencies. The RDA aim of "\textit{building the social and technical bridges that enable open sharing of data to achieve its vision of researchers and innovators openly sharing data across technologies, disciplines, and countries to address the grand challenges of society}" is at the core of this endeavour.

\subsection{RDA Groups and Outputs}

In October 2016, the RDA has a total of 72 Groups: 27 Working Groups (WG) and 45 Interest Groups (IG), with a significant increase from 15 Working Groups and 40 Interest Groups in October 2016. As explained in Genova (2016), the Working Groups are engaged in an 18-month effort to create "implementable" deliverables that will directly enable data sharing, exchange and interoperability. Interest Groups serve as a platform for communication and coordination among individuals, outside and within RDA, with shared interests in subjects related to data sharing. The Interest Groups also produce outputs such as surveys, recommendations, reports, and Working Group case statements. From the beginning of the RDA, Working Groups have been expected to develop implementable recommendations that the RDA would endorse. It appeared rapidly that these are not by far the only RDA products. The RDA now recognizes different kinds of outputs:

\begin{itemize}
	\item \underline{Recommendations} are the RDA equivalent of "specifications" or "standards" that other organisations recognize or endorse. The process is defined, with a Request for Comments (RfC) open to the whole membership and evaluation of adoption before the recommendation is endorsed.
	\item \underline{Supporting outputs} are WG or IG outputs which are not necessarily adoptable bridges. Upon request, they can go through a review or RfC, and if no major gap is identified they get the RDA brand.
	\item \underline{Other outputs} include workshop reports, published articles, survey results, etc., i.e., anything a WG or IG wants to register and report. Upon request, these are published and discoverable on the RDA website but have no level of endorsement.
\end{itemize}

It is worth noting that the quality of the RDA processes and outputs is being recognized by the European Multistakeholder Platform on ICT Standardisation (composed of EU member states and international and EU standards bodies representatives). Their evaluation of RDA as a non-standards body issuing technical specifications and RDA's first four recommendations as technical specifications to be used in public procurement in Europe was approved. When this paper is written, there is only a small step left before the European Commission Decision will be published in their Official Journal.

The first RDA Recommendations were very technical, on topics linked to Data Foundation and Terminology, PID Information Types, Data Type Registries, and Practical Policy (computer-actionable policy to deal with data). Since then a set of recommendations and outputs produced by Working Groups common to RDA and the ICSU WDS deals with data publishing, on publishing data bibliometrics, services, and workflows. The \textit{DSA} (Data Seal of Approval)\textit{-WDS Repository Audit and Certification Working Group} aligned the basic repository certification standards of the two organisations to produce a common framework for basic certification of trusted repositories.  A mechanism to reliably cite dynamic data is one of the recommendations which attracts the most attention. Other recommendations concern Metadata Standards Directory and Data Description Registry Interoperability. Finally one of the disciplinary groups, the Wheat Data Interoperability Working Group, produced a set of recommendations on Wheat Data Interoperability.

\section{RDA and Astronomy}

A variety of disciplines set up RDA Interest and Working Groups to work on their disciplinary interoperability frameworks. Astronomy successfully set up the International Virtual Observatory Alliance (IVOA) for that purpose, and thus does not need to propose a specific RDA Group. Other disciplines do. The Agriculture data community is an excellent example of using the RDA for disciplinary purposes. They decided from the very beginning of the RDA, in 2013, to use it as a global, neutral forum to work on their interoperability framework. They successively set up an Agriculture Data Interest Group, then a Wheat Data Interoperability Working Group, which defined the interoperability elements required by the International Wheat Initiative - one of RDA recommendations as explained above. They gather a global community, and now propose a Rice Data Interoperability Working Group on the model of the Wheat Data one, plus an Agrisemantics Working Group to tackle a more general topic, namely to gather community-based requirements and use cases for an infrastructure that supports appropriate use of semantics for data interoperability, with special focus on agriculture.

As explained in \citet{PO38_adassxxv}, astronomy representatives were present in RDA meetings and activities from its beginning. CDS has been a member of the series of European projects set up by the European Commission in support to RDA, and the organisational and sociological lessons accumulated when building the International Virtual Observatory have been used in the construction of the RDA organisation and processes. More generally, individuals involved in astronomical data sharing participate in or follow the activities of RDA Groups. The work of the RDA/WDS Groups which deal with data publication is of particular interest. The recommendation on dynamic data citation (keep track of queries) is also attracting interest from providers of highly dynamic resources, and is for instance being implemented in the framework of the Virtual Atomic and Molecular Data Centre VAMDC \citep{2016JMoSp.327..122Z}. The author also participated actively in the work of the DSA-WDS Repository Audit and Certification Working Group. CDS got both DSA and WDS Certification, several other astronomical data centres are also members of the WDS, and the IVOA is a WDS Network Member.

In Europe, the ASTERICS cluster project \citep{Pasian_adassxxv} \textit{Data Access, Discoverability and Interoperability} (DADI) Work Package, which ensures that the data from the large projects will be available in the Virtual Observatory Framework, has a subtask on liaison with the RDA and other generic data infrastructure projects such as EUDAT. This ensures some link between RDA and the large astronomy and astroparticle projects (CTA, E-ELT, EGO/VIRGO/ET, KM3Net, SKA) which participate in DADI.  One of the common topics is Provenance. DADI actively participates in the definition of the IVAO Provenance model, which expands the W3C provenance model to take into account our specific requirements, in particular those of CTA \citep{PO95_adassxxv}. This development is of high interest for the RDA Provenance Interest Group, which devoted a large fraction of its session at the Paris RDA Plenary (September 2015) to discuss it.

More generally, the IVOA is regularly informed of the RDA status and activities, through discussion in the Executive Board and in the Data Curation and Preservation Interest Group sessions, which include presentations from people who participate in the RDA activities or attend the Plenaries.

\section{Conclusion}

The RDA is a unique Forum for discussing all aspects of scientific data sharing with practitioners. Attending the RDA Plenaries is an excellent way to gather information on lessons learnt and best practices from a wide range of experts, both by participating in the Group sessions and through informal discussions. This is how for instance the CDS, which was already a member of the WDS, decided to apply also for the Data Seal of Approval, following discussions at the first RDA Plenary - and then actively participated in the RDA WG which aligned the two frameworks.

It is in the interest of astronomy to follow closely the RDA activities, in addition to using it as a forum to share best practices. This can be put in parallel with the evolution of the usage of IVOA Recommendations observed during the last years: they are now used by data providers, not only to build the interoperability layer on top of their data holdings \citep{arviset_adassxxv}, but also as building blocks inside their system: it saves time and effort to reuse useful developments prepared and validated by experts in an international context. The same is true for the RDA outputs, and it is worth for the discipline to identify those of interest for us.

\acknowledgements Fran\c{c}oise Genova acknowledges support from the Research Data Alliance - Europe 3 and ASTERICS Projects, funded by the European Commission (projects 653194 and 653477 respectively).

\bibliography{O5-3}  % For BibTex

\end{document}